\documentclass[aip,pof]{revtex4-1}
\usepackage{graphicx}
\draft 

\begin{document}


\title{On the role of non-uniform stratification and short-wave instabilities in three-layer quasi-geostrophic turbulence} 



\author{Gualtiero Badin}
\email[]{gualtiero.badin@uni-hamburg.de}
\affiliation{Institute of Oceanography, University of Hamburg, Bundesstrasse 53, 20146 Hamburg, Germany}


\date{\today}

\begin{abstract}
The role of short-wave instabilities on geostrophic turbulence is studied in a simplified model consisting of three layers in the quasi-geostrophic approximation. The linear stability analysis shows that short-wave instabilities are created by the interplay between the shear in the upper and the lower layers. If the stratification is non-uniform, in particular surface intensified, the linear growth rate is larger for short-wave instabilities than for long-wave instabilities and the layers are essentially decoupled, with the small scales growing independently. The fully developed homogeneous turbulence is studied in a number of numerical experiments. Results show that in both the case of equal layer depths and surface intensified stratification an inverse cascade in kinetic energy is observed. The modal kinetic energy spectra for the case with surface intensified stratification show higher energy for higher baroclinic numbers at small scales, due to the decoupling of the layers. As a result, while the case with equal layer depths shows large barotropic instabilities with large scale gradients of potential temperature, the surface intensified stratification is characterized by a transition from surface dynamics, characterized by a patchy distribution of vorticity, to interior dynamics, characterized by vorticity filamentation. The effect of the short-wave instabilities can be seen in the probability distribution functions of the potential vorticity anomaly, which reduces to a Gaussian distribution when the growth rate of the short-wave instabilities is larger than the growth rate for the long-wave instabilities. The surface intensified stratification also alters the vertical structure of the potential vorticity fluxes and shows deviations of the fluxes from a scaling obtained assuming that the turbulence acts as a downgradient diffusion. Experiments with a passive tracer shows a dominance of the coherent structures at large scales, and of filamentation at smaller scales, in the tracer dispersion.    
\end{abstract}


\maketitle 

\section{Introduction}
\label{sec:introduction}
The growth of disturbances in the atmosphere and in the ocean is well described by the classic baroclinic instability theory \cite{eady49,charney47}, which predicts that the available potential energy of a rotating, stratified fluid may be converted into the kinetic energy of the large-scale eddies. Focusing on the ocean, observations \cite{shcherbinaetal13} and high-resolution numerical simulations \cite{capetetal08I,kleinetal08} show however that the ocean is filled with dynamics at scales smaller than the mesoscale. These dynamics cover a wide range of different processes including internal waves \cite{garrettmunk72}, ageostrophic motions and frontogenetic processes \cite{hoskinsbretherton72}. Even if at small scales a number of factors might interplay in the breakdown of the quasi-geostrophic (QG) approximation, such as the existence of dynamics at $O(1)$ Rossby number or the spontaneous generation of waves by loss of balance (for reviews on these processes see Thomas et al \cite{thomasetal08} and Vanneste \cite{vanneste13}), this study will concentrate on the generation of instabilities at scales smaller than the deformation radius even using the QG approximation. 

Different mathematical models have been proposed to describe the emergence of short-wave instabilities in balanced geostrophic flows. While the classic two-layer QG model for baroclinic instability \cite{phillips54} is characterized by a cut-off at high wavenumbers, a form of two-layer QG baroclinic instabilities able to produce short-wave instabilities separate from long-wave instabilities was proposed by Blumen \cite{blumen79}, who studied the effects of having two layers with constant shear overlying each other, essentially resulting in the coupling of two Eady modes. However, results of linear stability show that, at least in the parameter range explored by Blumen \cite{blumen79}, the growth rate associated with short-wave instabilities is much smaller than the one associated with long-wave instabilities. A QG model that is characterized by the growth of short-wave instabilities is the classic model by Charney \cite{charney47}, which does not possess a short-wave cut-off but has instead Green modes \cite{green60} growing at high wavenumbers. The stability analysis of the Charney model requires an interaction between the vertical shear (the horizontal buoyancy gradient) at the surface and the interior potential vorticity (PV) gradient. Ocean currents characterized by positive vertical shear at the surface and a positive PV gradient in the interior will thus result in a stable profile. In order to have instabilities, a change in the PV gradient in the water column or a shear reversal close to the surface, which can be realized by the presence of a surface mixed layer \cite{tullochetal11}, are thus required. A model that has received much attention for its capability to create short-wave instabilities is the Surface Quasi-Geostrophic (SQG) model \cite{blumen78,heldetal95}. SQG dynamics is characterised by two invariants, total energy and an active surface buoyancy, and possess an inverse cascade of total energy at low wavenumbers and a direct cascade of potential temperature (or density) variance at the boundary (the surface) for high wavenumbers. The Charney and the SQG model are characterised, respectively, by a finite depth of penetration of the short-wave instabilities from the boundary, also known as the Charney depth, and at an e-folding depth dependent on the wavenumber. SQG dynamics possess also both mathematical and physical analogies with the three dimensional Euler equation for incompressible flows \cite{constantinetal94} making it a simpler paradigm to study the formation of singularities.    

This study will concentrate on the mechanism of how to generate geostrophic instabilities at scales smaller than the deformation radius using a three-layer QG model, and on the characterisation of the resulting turbulence in the presence of these instabilities. This will allow to characterize the role of short-wave instabilities using the QG approximation and to make a qualitative comparison with the knowledge of turbulence created for example by ageostrophic processes. The three-layer QG model retains a mathematical simplicity that allows for the control of the parameters, such as the relative shear between the different layers, setting the physics of the system. It is also less expensive computationally to run than a multilayer QG model, required for example to resolve the Green modes in a Charney model. Finally, even if they are both based on the QG approximation, it has a different physics than the SQG approximation, which does not allow for internal PV gradients and that allows, instead, for frontogenesis at the surface. 
The three-layer model shows linear growth both at low and high wavenumbers \cite{davey77}. With three-layers it is possible to find intervals of relative shear between the upper and lower layers in which the system has exponential growth at both low and high wavenumbers. An interpretation of the emergence of short-wave instabilities in the three-layer problem was given by Davey \cite{davey77}, making use of the baroclinic instability analysis made by Bretherton \cite{bretherton66}. In a three-layer model with equal layer depths, short-wave instabilities can be excited for weak shear between the middle and lower
layers due to due to the large potential vorticity gradient in the lower layer. The layers are however decoupled at short wavelengths. The interface between the two uppermost layers thus acts as a boundary and the system behaves like a two-layer system over a constant topographic slope, with the short-wave instabilities confined in the lower layer. If the stratification is uniform, i.e. the layers have equal depth, the linear stability analysis shows that the growth of short-wave instabilities is weaker than that of the long-wave instabilities.   

However, if the stratification is not uniform and the flow is surface intensiﬁed, at sufficiently strong shear the maximum growth rate can change from long waves associated with the upper layer to short waves associated with the lower layer \cite{smeed88,benilov95}. The generation of short wave baroclinic instabilities in presence of surface intensified flows is reminiscent of the creation of small-scale instabilities by SQG flows and deserves to be explored.

While a number of studies were dedicated to the analytical properties of the linear and nonlinear stability of the three-layer model \cite{ikeda83,vanneste95,paretvanneste96,pichevin98,smithvallis99,olascoaga01,sutyrin07,badinetal09}, very few studies explored the fully developed turbulent regime. Held and O'Brien \cite{heldobrien92} studied the vertical structure of the eddy PV fluxes, exploring the role of the mean shear and of the curvature of the flow, and investigated if the PV fluxes are controlled by the most unstable modes or by a baroclinic adjustment process. The interest arises because while in the two-layer model the knowledge of the PV eddy fluxes in a layer determines the complete information on the system due to the fact that the fluxes must integrate vertically to zero, in three or more layers the fluxes might be distributed differently, even if they must still vertically integrate to zero. Held and O'Brien \cite{heldobrien92} did not put emphasis however on the role of short-wave instabilities that, being confined to the lower interface, might introduce a vertical structure in the fluxes.

In Section \ref{sec:linear} we will investigate the role of equal layer depths and surface intensified stratification in the linear stability of a three-layer problem. In Section \ref{sec:turbulence} we will study the properties of homogeneous QG turbulence in the presence of short-wave instabilities. The effects of short-wave instabilities on the kinetic energy (KE) spectra, the probability density function (PDFs) of PV and the vertical structure of the eddy PV fluxes as well as a qualitative application to the lateral stirring of a passive tracer will be studied. Finally in Section \ref{sec:discussion} the results of the three-layer model with surface intensified stratification will be discussed to make a qualitative comparison with the results from SQG dynamics that are present in the literature.

\section{Linear stability analysis}
\label{sec:linear}
\subsection{Formulation of the problem}
\label{sec:theory}
Consider a three-layer system in the QG approximation. The basic flow given by the zonal velocities $U_{i},~i=1,2,3$, which are constant within each layer of depth $H_{i}$. The linearized non-dimensional equations for the PV anomaly can be written as
\begin{equation}
\frac{D}{Dt_{i}} \left[ \nabla^{2} \psi_{i} + \sum_{j=1}^{3} A_{i,j} \psi_{j} \right] + \frac{\partial \psi_{i}}{\partial x} \frac{\partial \Pi_{i}}{\partial y} + \delta_{i,3} r_{EK} \nabla^{2} \psi_{i} = 0, 
\label{eq:pv}
\end{equation}
where $\psi_{i}$ are the perturbation streamfunctions in each layer, 
\begin{equation}
q_{i}=\nabla^{2} \psi_{i} + \sum_{j=1}^{3} A_{i,j} \psi_{j} ,
\label{eq:q}
\end{equation}
is the perturbation PV, and
\begin{equation}
A =
 \left( \begin{array}{ccc}
-F_{1} & F_{1} & 0 \\
F_{21} & -F_{21}-F_{23} & F_{23} \\
0 & F_{3} & -F_{3} \end{array} \right) ,
\label{eq:A}
\end{equation}
where
\begin{equation}
\begin{array}{cc}
F_{1}=\frac{f_{0}^{2} L^{2}}{(2 \pi)^{2} g'_{1} H_{1}}, & F_{3}=\frac{f_{0}^{2} L^{2}}{(2 \pi)^{2} g'_{2} H_{3}}, \\
F_{21}=\frac{f^{2}_{0} L^{2}}{(2 \pi)^{2} g'_{1} H_{2}}, & F_{23}=\frac{f_{0}^{2} L^{2}}{(2 \pi)^{2} g'_{2} H_{2}}, 
\end{array}
\label{eq:F}
\end{equation}
are the internal Froude numbers where $f_{0}$ is the Coriolis frequency, $L$ is the typical horizontal scale of the flow and the $g'_{i}$ are the reduced gravities of the system. 
\begin{equation}
\frac{\partial \Pi_{i}}{\partial y} = \beta+\sum_{j=1}^{3} A_{i,j} U_{j} ,   
\label{eq:pvgrad}
\end{equation}
is the basic state PV gradient, where $\beta$ is the non-dimensional meridional gradient of the Coriolis frequency. 
\begin{equation}
\frac{D}{D t_{i}} = \frac{\partial}{\partial t} + U_{i} \frac{\partial}{\partial x} , 
\label{eq:lagrangian}
\end{equation} 
are the Lagrangian derivatives. Finally, the last term of (\ref{eq:pv}) is an Ekman damping, where $r_{EK}$ is the Ekman friction coefficient and $\delta_{i,3}$ is the Kroenecker delta.

Defining $\epsilon=H_{1}/H_{3}=H_{2}/H_{3}$, it is apparent that if $\epsilon = 1$, the system corresponds to the case of equal layer depths, while if $\epsilon<<1$ the system corresponds to the case with a deep bottom layer with non-dimensional depth $H_{3}=H$ overlied by two thin layers of equal nondimensional depth. Under these assumptions, corresponding to the case of surface intensified stratification, and assuming equal reduced gravities at the two interfaces, one has $F_{21}=F_{23}=F_{1}$. Setting, for simplicity, $U_{2}=1$ and $U_{3}=0$, and assuming that the flow is confined in a double periodic domain in the horizontal and with rigid lid and flat bottom boundary conditions in the vertical, solutions can be sought in the form of normal modes
 \begin{equation}
\psi_{i}=\phi_{i} \exp{\left[ i(kx+ly-\sigma t) \right]} ,  
\label{eq:nm}
\end{equation} 
where $k,l$ are the zonal and meridional wavenumbers, respectively, and $\sigma$ is the wave frequency, that is related to the wave propagation speed in the zonal direction by $c=\sigma / k$. Substitution of (\ref{eq:nm}) in (\ref{eq:pv}) gives the eigenvalue problem for the amplitudes 
 \begin{equation}
det~D=0 ,  
\label{eq:nmprob}
\end{equation} 
where
\begin{widetext}
\begin{equation}
D =
 \left( \begin{array}{ccc}
c'\left( K^{2} + \frac{1}{\epsilon H}\right) +K^{2}-\frac{\beta}{S} & -\frac{1}{\epsilon H}(c'+1) & 0 \\
-\frac{c'}{\epsilon H} & c' \left( K^{2} + \frac{2}{\epsilon H} \right) + \frac{1}{\epsilon H} \left( 1 - \frac{1}{S} \right) - \frac{\beta}{S} & -\frac{c'}{\epsilon H} \\
0 & -\frac{1}{H}\left( c' - \frac{1}{S} \right) & c' \left( K^{2} + \frac{1}{H} \right) -\frac{K^2}{S} - \frac{\beta}{S} -i r_{EK} \frac{K^{2}}{kS} \end{array} \right) ,
\label{eq:D}
\end{equation} 
\end{widetext}
where we have used the notation by Smeed \cite{smeed88} and we have defined $K^{2}=\frac{g'}{f^{2} L^{2}} \left( k^{2} + l^{2} \right)$, $c'=(1 - c) / U_{1}$ and $S=U_{1} - 1$. The last quantity, $S$, is the relative shear (in general form, $S=(U_{1} - U_{2}) / (U_{2} - U_{3})$), that will play an important role in the rest of the study. It can be seen that for $U_{1}>U_{2} \Rightarrow S>0$, $U_{1}=U_{2} \Rightarrow S=0$ and $U_{1}<U_{2} \Rightarrow S<0$, with the particular case $U_{1}<0<U_{2} \Rightarrow S<-1$. In an oceanographic context, $S<0$ has a physical meaning in the presence of counter currents below the surface. 

Equation (\ref{eq:nmprob}) results in a cubic equation for $c'$. The three solutions of the cubic can be either all be real or one real and two complex conjugates. While the real solutions correspond to neutral waves, the complex conjugates correspond to exponential modes, with high-wavenumber instabilities associated with the lower layer \cite{davey77}. Approximated analytical solutions were found by Smeed \cite{smeed88} for the cases with $\beta=0$ and with two deep layers overlying a thin lower layer (which, in the absence of bottom friction, is symmetric to the case of a thin upper layer over two deep bottom layers). Using an asymptotic expansion with the ratio between the depth of the thin layer and the deep layers as small parameter, Smeed \cite{smeed88} was able to find that short-wave instabilities are possible for $S<1$. Benilov \cite{benilov95} studied the case of two thin upper layers, as introduced here, small $\beta$ and small mean flow velocities in the thin layers. Results were derived analytically for medium and long wavenumbers. In the following we explore the generation of short wavenumber linear instabilities for the same stratification used by Benilov \cite{benilov95}, in order to study the effects of surface intensified stratification. While analytical solutions using the QG approximation are possible for short-wave instabilities, we will study the more general case in which $\beta$ and the mean flow velocities are not small. 

\subsection{Results of the linear stability analysis}
\label{sec:linearresults}
All calculations here presented were done with $F_{0}=\frac{f_{0}^{2} L^{2}}{(2 \pi)^2 g' H_{0}}$ fixed to the value of $F_{0}=24$, where $H_{0}$ is the typical value of the dimensional total depth of the domain. With $H_{0}=10^{3}$m, the density difference between layers fixed to $\Delta \rho=0.1$, and $g'=10^{-3}$m s${}^{-2}$, this choice for $F_{0}$ gives a value of $L \approx 3 \times 10^5$m. While this value is probably unrealistically large for an oceanic mesoscale eddy, it allows for a wider range of unstable wavenumbers. In the following, results for $\beta=1$ and $r_{EK}=0.5$ are presented. The cases with equal layer depths and surface intensified stratification are compared. In the case with uniform stratification the nondimensional equal-layer depths are set to $H_{i}=1/3$. The resulting wavenumber corresponding to the deformation radius is $k_{d}=8.49$. In the case with surface intensified stratification the layer depths are set to $H=0.9$, $\epsilon H=0.05$. The corresponding deformation wavenumber is $k_{d}=14.23$. 
\subsubsection{Equal layer depths}
\begin{figure}[t]
  \noindent\includegraphics[width=30pc,angle=0]{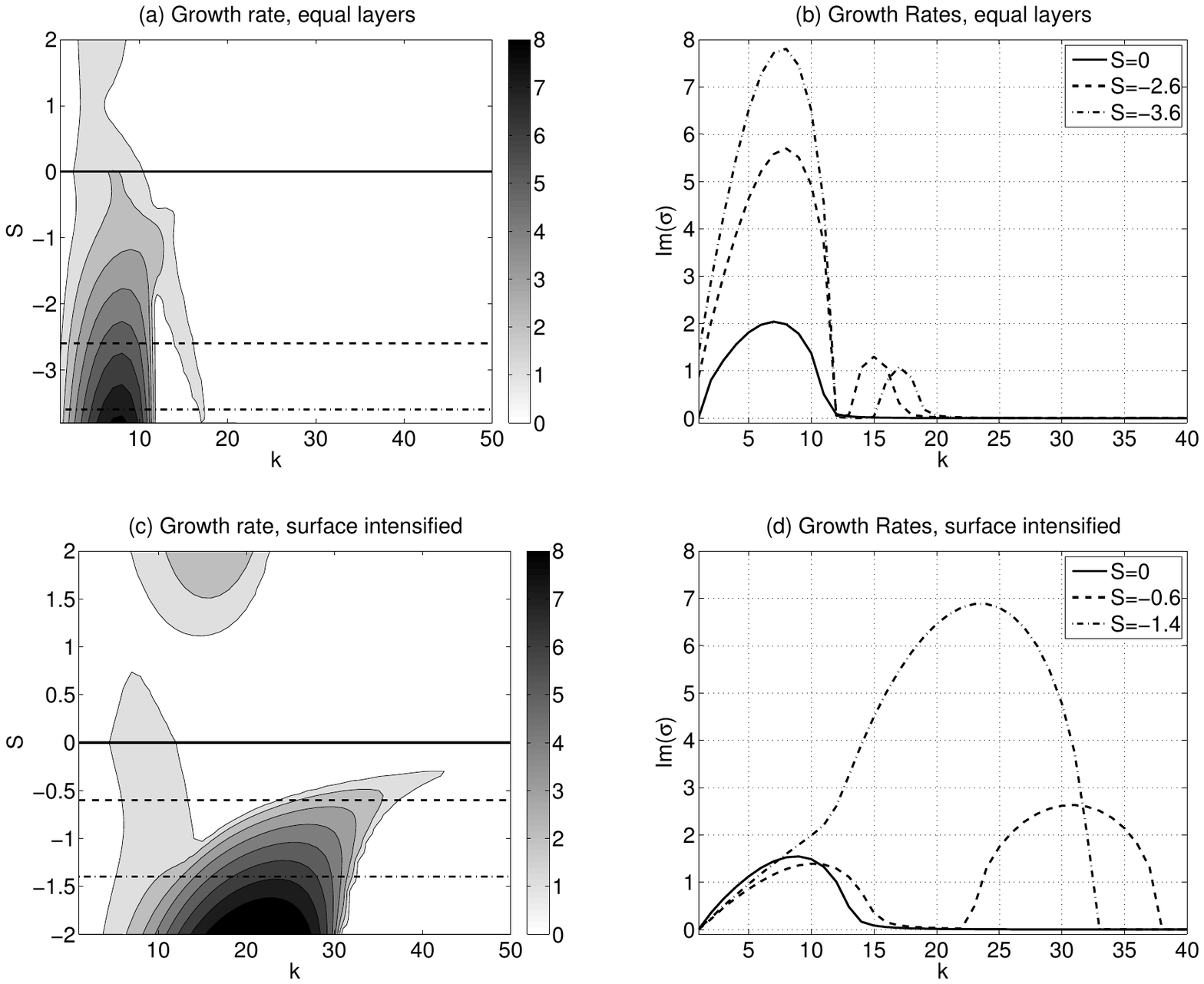}\\
  \caption{Left panels: growth rates as a function of the zonal wavenumber $k$ for $l=0$ and of the relative shear $S$ for the cases with (a) equal layer depths and (c) surface intensified stratification. Right panels: growth rates for $l=0$ for (b) equal layer depths along $S=0$ (full line), $S=-2.6$ (dashed line) and $S=-3.6$ (dot-dashed line) and (d) surface intensified stratification along $S=0$ (full line), $S=-0.6$ (dashed line) and $S=-1.4$ (dot-dashed line).}\label{f1}
\end{figure}

The exponential growth rate for the case with equal layer depths shows a low-wavenumber peak in the growth rates confined within $0 \le k \le 17$ (Figure \ref{f1}a,b). For $S<1$, as $S$ decreases the maximum growth rate increases. Between $-0.45 \le S \le -0.35$ and for $S \le -1.9$, a second peak of unstable modes appears at higher wavenumbers. For $S<-1.9$, the distance in wavenumber space between the main peak at low wavenumbers and the secondary peak at high wavenumbers increases as $S$ decreases. For $S<-3.9$, the short-wave instabilities disappear. It should be noted that the secondary peak in the growth rate is always smaller than the peak for long waves. 

The dependence of the maximum value of the growth rate on $S$ and on $\beta$ shows a linear increase as $S$ decreases for $S<-1$, while it flattens for values of $S$ close to zero (Figure \ref{f1b}a). For these values of $S$, a small difference in the growth rates for different values of $\beta$ appears, with lower values corresponding to higher values of $\beta$, due to the low wavenumber cutoff introduced by the planetary vorticity gradient. For $S<-2.2$, different values of $\beta$ correspond to the same maximum value of the growth rate of the instabilities. 

Finally, the dependence of the wavenumber $k$ corresponding to the maximum value of the growth rate on $S$ and on $\beta$ shows that for $S>-2.2$, $k$ varies between $k=7$ and $k=10$, with higher values assumed by the cases with higher $\beta$ (Figure \ref{f1b}b). This effect can be explained by the fact that the $\beta$ effect decreases the growth rates for lower wavenumbers, moving the maximum, which will have a lower value than for lower values of $\beta$, at higher wavenumbers. For $S<-2.2$, the value of $k$ does not change from the value $k=8$.

\subsubsection{Surface intensified stratification}
\begin{figure}[t]
  \noindent\includegraphics[width=30pc,angle=0]{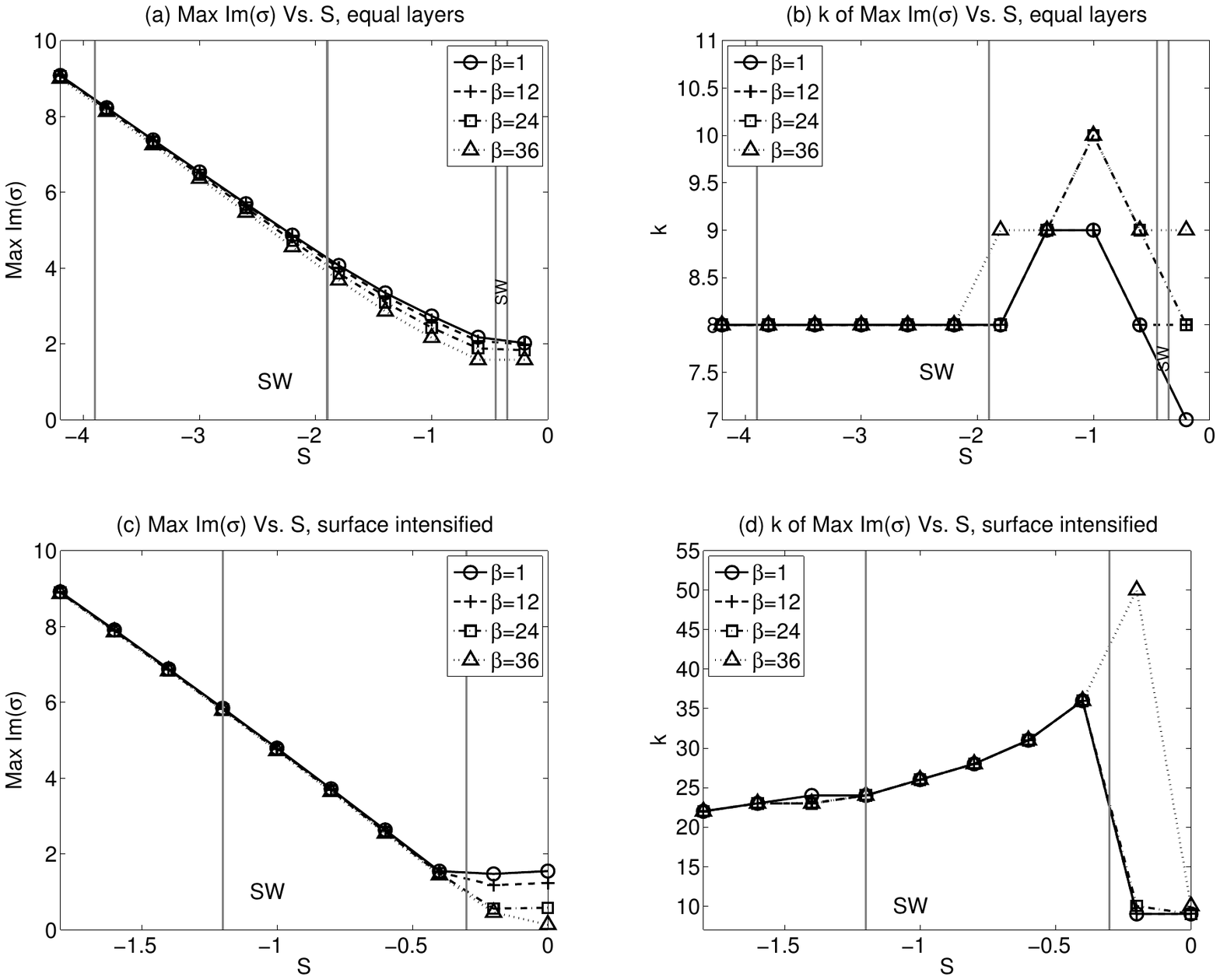}\\
  \caption{Left panels: maximum growth rate as a function of $S$ for (a) equal layer depths and (c) surface intensified stratification. Right panels: zonal wavenumber correspondent to the maximum growth rate as a function of $S$ for (b) equal layer depths and (d) surface intensified stratification, for $\beta=1$ (circles, full lines), $\beta=12$ (crosses, dashed lines), $\beta=24$ (squares, dot-dashed lines) and $\beta=36$ (triangles, dotted lines). Vertical gray lines indicate the limits of the $S$ intervals inside which short-waves instabilities are present.}\label{f1b}
\end{figure}

The case with surface intensified stratification shows that for $-0.3 < S \le 0$ the growth rate has a maximum at low wavenumbers (Figure \ref{f1}c, and \ref{f1b}c,d). In the interval $-1.2 \le S \le -0.3$, a second peak of unstable modes appears at high wavenumbers. Differently from the case with equal layer depths, the peak at high wavenumbers is larger than the peak at low wavenumbers. A second major difference from the case with equal layer depths lies in the larger separation between the two peaks and in the extension of the unstable modes to high wavenumbers, reaching $k \le 50$. As $S$ decreases, the peak at high wavenumbers moves toward lower wavenumbers. For $S<-1.2$, the peak at high wavenumbers merges with the peak at lower wavenumbers. 

Increasing the value of $\beta$ decreases the amplitude of the growth of long waves, leaving unaltered the growth of short waves. This results in the fact that for $S \le -0.4$, the value the corresponds to the onset of short wave instabilities, the maximum growth rate increases linearly as $S$ decreases and does not show dependence on $\beta$ (Figure \ref{f1b}c). For higher values of $S$, the maximum growth rate flattens and shows a dependence on $\beta$, with higher values for smaller $\beta$.

The dependence of the zonal wavenumber corresponding to the maximum growth rate on $S$ and on $\beta$ reflects the emergence of the higher peak at high wavenumbers for $S \le -0.2$, and the continuous transition to lower wavenumber as $S$ decreases (Figure \ref{f1b}d). The transition to high wavenumbers of the zonal wavenumber corresponding to the maximum growth rate does not show dependence on the value of $\beta$ except for the case with $\beta=36$, for which the the growth of long-wave instabilities is close to zero, resulting in a transition to high wavenumbers for higher values of $S$.

\section{Fully developed turbulence}
\label{sec:turbulence}
To study the effects of short-wave instabilities in the fully developed QG turbulence, a number of numerical simulations were performed using a dealiased spectral QG model with $512~\times~512$ equally spaced horizontal grid-points, rigid lid and flat bottom. The numerical model is the same previously employed by Smith and Vallis \cite{smithvallis01,smithvallis02}. The domain is doubly-periodic in the horizontal in the interval $0 \le x,y < 2 \pi$. Simulations with $F_{0}=24$ and $F_{0}=1$ have been performed for comparison. With the number of vertical layers fixed to 3, two sets of numerical simulations were performed with equal layer depths and surface intensified stratification using the same parameters stated in the linear stability analysis. Simulations with values for the relative shear $-4 \le S \le 2$ were performed. While all simulations were performed using the reference value $\beta=1$, a number of sensitivity experiments with increased $\beta$ were done. A third set of experiments was performed using equal layer depths but with different reduced gravity  between the upper and the lower two-layers. The model was initially forced at wavenumber $k=10$, which is very close to the wavenumber of the maximum growth rate from linear stability analysis. The initial energy has Gaussian distribution with a half-width of 2 and a maximum value of $10^{-3}$. An exponential enstrophy filter is applied, with the cut-off wavenumber set to $k=205$. The model is run until a statistical steady state is achieved.    

\subsection{Results for fully developed turbulence}
\label{sec:results}

\subsubsection{Kinetic energy spectra}
\label{sec:spectra}
\begin{figure}[t]
  \noindent\includegraphics[width=30pc,angle=0]{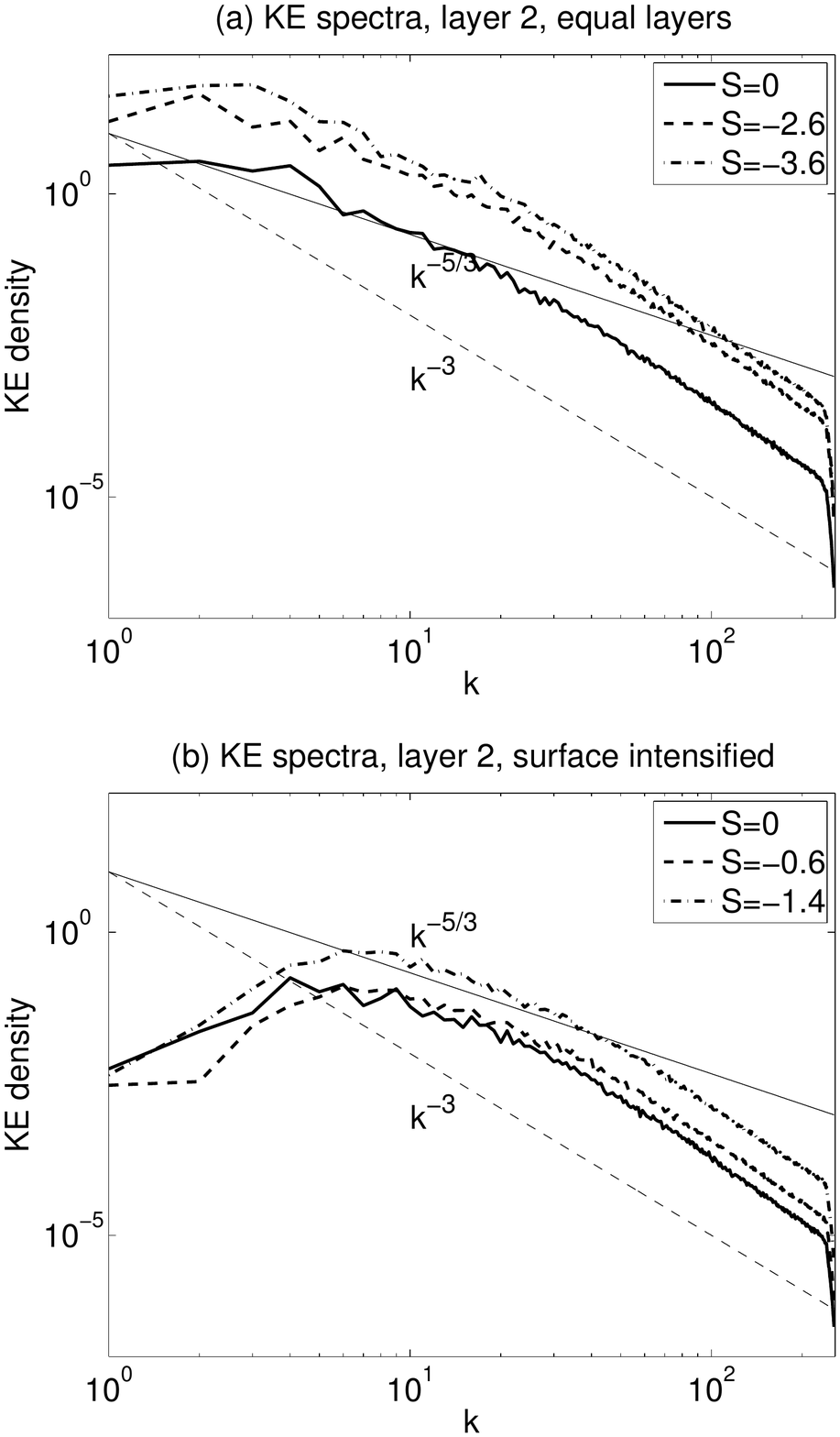}\\
  \caption{(a) Kinetic energy spectra for layer two for equal layer depths and $S=0$ (full line), $S=-2.6$ (dashed line) and $S=-3.6$ (dot-dashed line). (b) Kinetic energy spectra for layer two for surface intensified stratification and $S=0$ (full line), $S=-0.6$ (dashed line) and $S=-1.4$ (dot-dashed line). Lines with constant slopes $-3$ and $-5/3$ are added for comparison.}\label{f4a}
\end{figure}

The KE spectra in layer two (Figure \ref{f4a}) show an inverse cascade of energy with slope $k^{-5/3}$ at scales larger than the deformation radius and a forward cascade with $k^{-3}$ at scales smaller than the deformation radius. The cases with equal layer depths and surface intensified stratification differ for a larger spectral energy associated with the case with equal layer depths. The case with equal layer depths also shows a larger spectral energy difference at all wavenumbers for different $S$ than the case with surface intensified stratification. At larger scales, the spectra show a flattening due to saturation of the domain.

In order to have information on how much energy is contained in each vertical mode, i.e. in the barotropic and in the first and second baroclinic modes, a modal energy budget can be derived from the modal PV equation \cite{flierl78,fuflierl80}.
To do so, first rewrite (\ref{eq:nm}) as
 \begin{equation}
\psi_{i}=\phi^{m}_{i}(t) \exp{\left[ i(kx+ly) \right]} ,  
\label{eq:mnm}
\end{equation} 
where $\phi^{m}_{i}$ represents the amplitude of the vertical mode $m$ in the layer $i$. In (\ref{eq:mnm}), $m=0,1,2$, which indicate, respecively, the barotropic, first baroclinic and second baroclinic modes. $\phi^{m}_{i}$ can be further split in a time-dependent and a vertical structure components
 \begin{equation}
\phi^{m}_{i}(t)=\alpha (t) F^{m}_{i} .  
\label{eq:mnm2}
\end{equation} 
Insertion of (\ref{eq:mnm2}) in the linearized version of (\ref{eq:pv}) and separation of variables yields the Sturm-Liouville problem
 \begin{equation}
\sum_{j=1}^{3} A_{i,j} F^{m}_{j}=-\lambda^{m} F^{m}_{i} ,  
\label{eq:msl}
\end{equation}
which can be solved using rigid rid and flat bottom boundary conditions. The resulting modes are normalized as
 \begin{equation}
\sum_{j=1}^{3} F^{m}_{j} F^{n}_{j}= \delta_{m,n} .  
\label{eq:mnorm}
\end{equation}   
Insertion of (\ref{eq:mnm2}) into the full nonlinear version of (\ref{eq:pv}) and summation on all the vertical layers gives 
\begin{equation}
\frac{\partial Q^{m}}{\partial t}+ \sum_{p,s} \epsilon^{mps} \hat{J}(\alpha^{p},Q^{s}) + i k \beta \alpha^{m} = - r_{EK} K^{2} \alpha^{m}   
\label{eq:mpv}
\end{equation}  
where $\hat{J}$ in the spectral Jacobian and
\begin{equation}
Q^{m}=-\left( K^{2} + \left(\lambda^{m} \right)^{2} \right) \alpha^{m} .   
\label{eq:mQ}
\end{equation}
In (\ref{eq:mpv}) 
\begin{equation}
\epsilon^{mps} = \sum_{i=1}^{3} F_{i}^{m} F_{i}^{p} F_{i}^{s} H_{i} ,   
\label{eq:mtrip}
\end{equation}
is the triple interaction coefficient. Multiplication of (\ref{eq:mQ}) by $-\alpha^{m~*}$, where the star indicates the complex conjugate, yields the equation for the total energy in each vertical mode $m$, $E^{m}$
\begin{equation}
\frac{d E^{m}}{dt}  = T^{m} + D^{m} ,   
\label{eq:mtrip}
\end{equation}
where $T^{m}$ represents the internal energy transfer by triple interactions in each vertical mode $m$ and $D^{m}$ represents the sink of energy due to Ekman dissipation in each vertical mode $m$.

\begin{figure}[t]
  \noindent\includegraphics[width=30pc,angle=0]{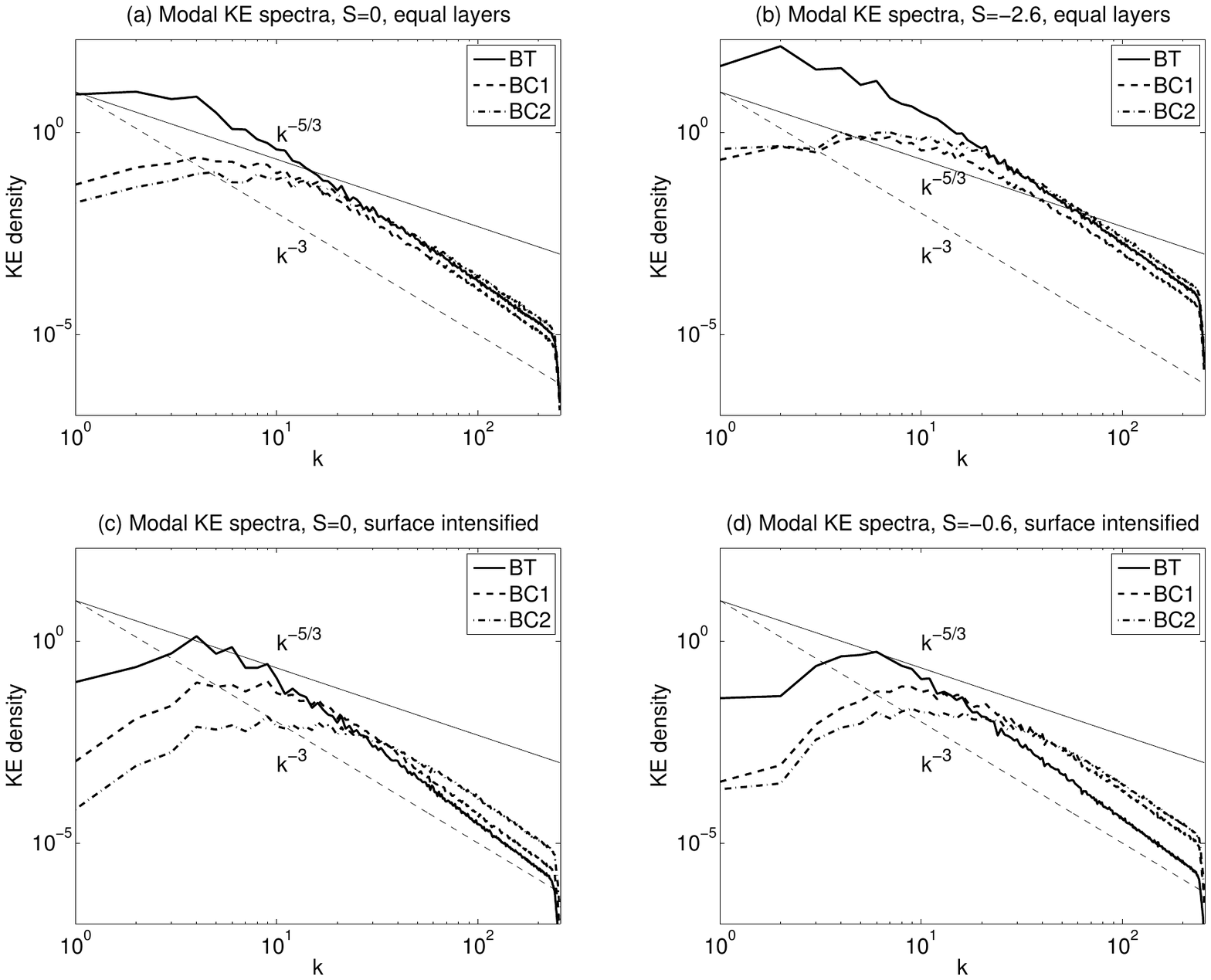}\\
  \caption{Modal kinetic energy spectra for (a) equal layer depths and $S=0$, (b) equal layer depths and $S=-2.6$, (c) surface intensified stratification and $S=0$ and (d) surface intensified stratification and $S=-0.6$. Full line: barotropic mode, dashed line: first baroclinic mode, dot-dashed line: second baroclinic mode. Lines with constant slopes $-3$ and $-5/3$ are added for comparison.}\label{f4}
\end{figure}

The modal KE spectra for the fully turbulent regimes are shown in Figure \ref{f4}. The total energy (i.e. integrated over all the wavenumbers) in each vertical mode (not shown) shows that the energy in the barotropic mode is higher than the energy in the baroclinic modes for all the cases considered, however details change significantly between the case with equal layer depths and the case with surface intensified stratification. In particular, the case with equal layer depths has higher energy in the barotropic mode than the case with surface intensified stratification, reflecting the inhibition of the inverse energy cascade toward the barotropic mode by surface intensified stratification in QG turbulence \cite{smithvallis01,smithvallis02}.

In detail, the case with equal layer depths shows that for $S=0$, the first baroclinic mode (dashed line) has lower energy than the barotropic (full line) and the second baroclinic mode (dot-dashed line) at all scales (Figure \ref{f4}a). The barotropic mode dominates the second baroclinic mode for $k \le 22$, while the second baroclinic mode shows slightly higher energy levels for $k \ge 63$. For $S=-2.6$, in the presence of short-wave instabilities, the general picture is unchanged, apart from the fact that the barotropic mode shows larger KE than for the case $S=0$ (Figure \ref{f4}b).

In particular, the case with surface intensified stratification shows instead that, for scales smaller than a certain cut-off, both baroclinic modes dominate the barotropic mode. In detail, for $S=0$ (Figure \ref{f4}c), for $k \ge 15$ the first baroclinic mode has more energy than the barotropic mode, while the second baroclinic mode has more energy than the first baroclinic mode for $k \ge 25$. At smaller scales, corresponding to $k \ge 37$, the energy contained in the second baroclinic mode is larger than the energy in the first baroclinic mode. For $S=-0.6$ (Figure \ref{f4}d), the energy in the second baroclinic mode is closer to the energy of the first baroclinic mode for all $k$. 

\subsubsection{Streamfunction, potential temperature and PV anomaly fields}
\label{sec:fields}
\begin{figure}[t]
  \noindent\includegraphics[width=30pc,angle=0]{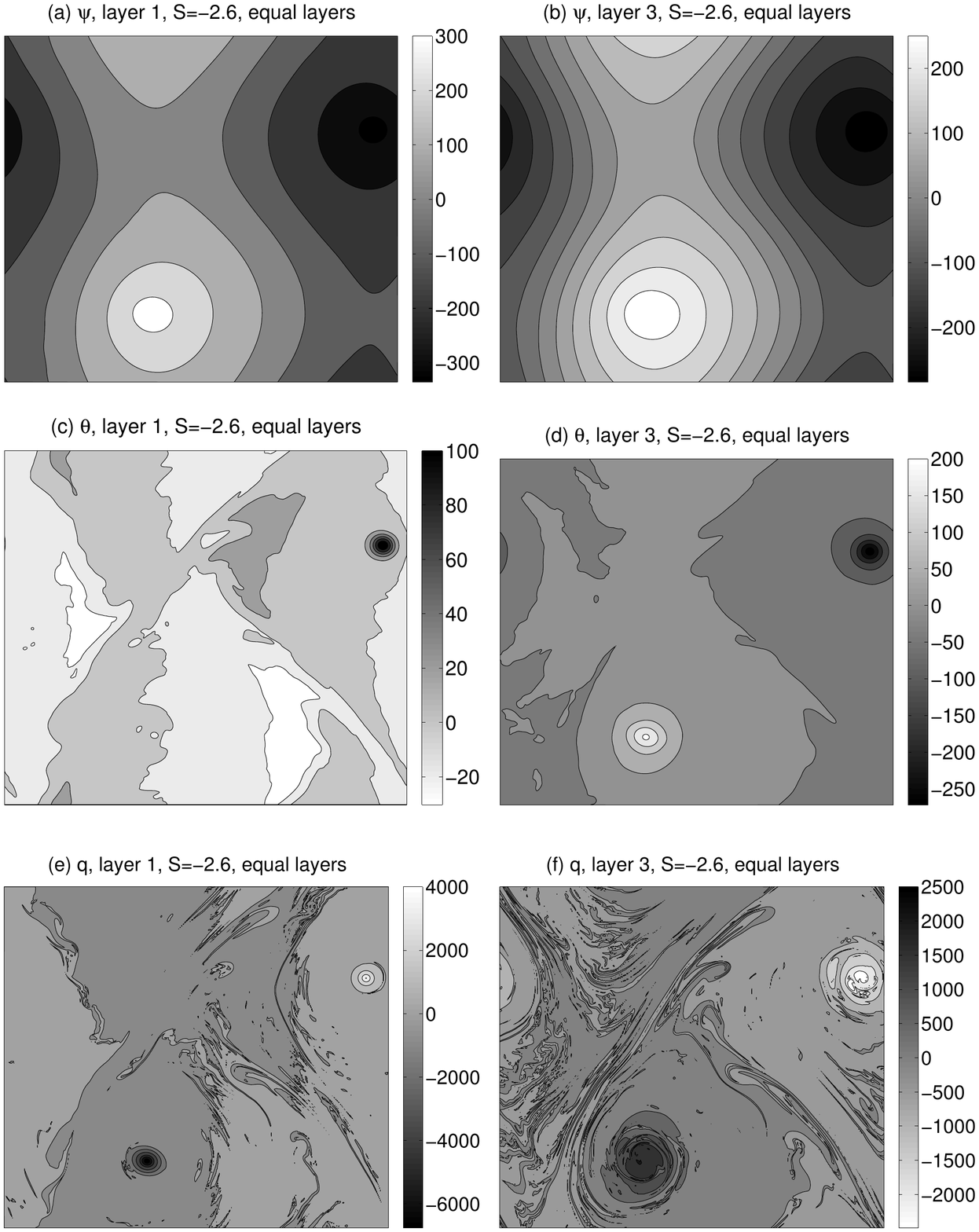}\\
  \caption{Snapshots of (a,b) streamfunction $\psi$, (c,d) potential temperature $\theta=\psi_{z}$, (e,f) potential vorticity anomaly $q$ for layer 1 (left panels) and layer 3 (right panels) for equal layer depths and $S=-2.6$.}\label{f2}
\end{figure}

\begin{figure}[t]
  \noindent\includegraphics[width=30pc,angle=0]{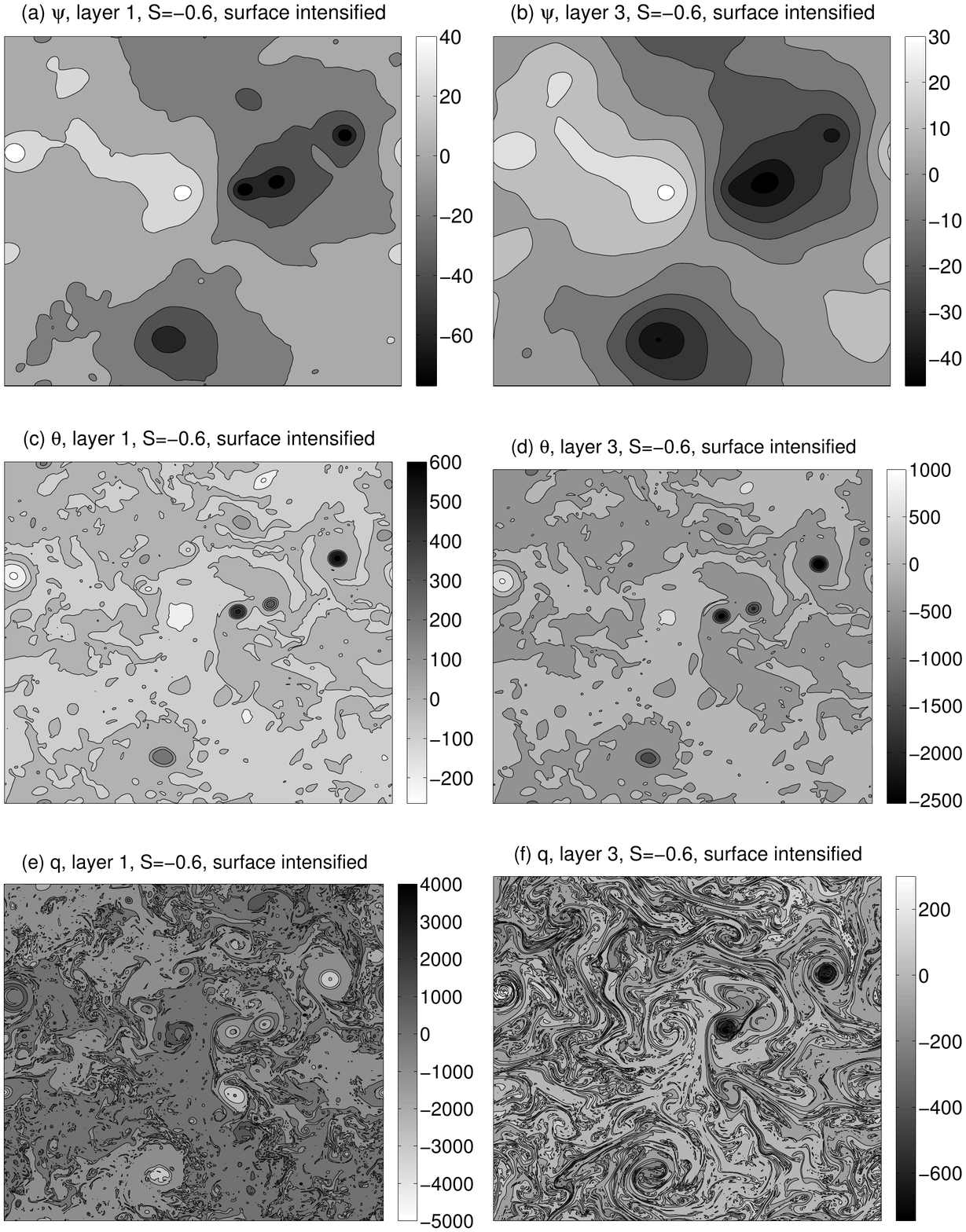}\\
  \caption{Snapshots of (a,b) streamfunction $\psi$, (c,d) potential temperature $\theta=\psi_{z}$, (e,f) potential vorticity anomaly $q$ for layer 1 (left panels) and layer 3 (right panels) for surface intensified stratification and $S=-0.6$.}\label{f3}
\end{figure}

The shape of the KE spectra is reflected on the appearance of the flow field and the associated potential temperature and PV anomaly fields. Snapshots of the streamfunction $\psi$, potential temperature $\theta=\partial \psi / \partial z$ and PV anomaly $q$ for the case with equal layer depths and surface intensified stratification are shown in Figure \ref{f2} and \ref{f3} respectively, for the upper layer (left columns) and lower layer (right columns). The snapshots are shown for values of the relative shear correspondent to the exponential growth of short-wave instabilities. 

The case with equal layer depths and $S=-2.6$ shows large, barotropic localized coherent structures, with intense cores (Figure \ref{f2}). In detail, the potential temperature field (Figure \ref{f2}c,d) shows the presence of both strong gradients in small regions surrounding the cores of the coherent structures, and weaker gradients aligned with the saddle-shaped geometry of the flow at all depths (Figure \ref{f2}a,b). The PV anomaly field follows the potential temperature field, with signatures of secondary roll-up instabilities (Figure \ref{f2}e,f).

The case with surface intensified stratification shows instead a different picture (Figure \ref{f3}). The potential temperature field shows a lack of the large scale gradients but has instead a patchy distribution around the signatures of intense vortex cores (Figure \ref{f3}c,d). The distribution of $\theta$ does not seem to show depth dependence. The PV anomaly field shows instead variability with depth (Figure \ref{f3}e,f). In the upper layer the PV anomaly shows the sign of localized coherent structures, with a transition to filamentary structures at depth that reminds of a passive tracer in the Batchelor regime. Interestingly this transition happens without a transition to a shallower slope in the spectra, that would require instead higher vertical resolution to be seen \cite{tullochsmith09}.  

\begin{figure}[t]
  \noindent\includegraphics[width=30pc,angle=0]{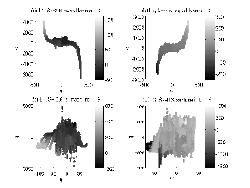}\\
  \caption{Relationship between streamfunction $\psi$, PV and potential temperature $\theta$ for equal layer depths and $S=-2.6$ in (a) layer 1 and (b) layer 3, and for surface intensified stratification and $S=-0.6$ in (c) layer 1 and (d) layer 3.}\label{f3a}
\end{figure}

The behavior of the different fields in the cases with equal layer depths and surface intensified stratification is clearly visible in the $\psi / q$ scatter plots. In the case with equal layer depths and $S=-2.6$ the dominance of the flow by large, barotropized coherent structures induces a "sinh-like" appearance to the $\psi / q$ relationship (Figure \ref{f3a}a,b), as observed in different asymptotic configurations of freely-decaying and forced quasi-two-dimensional turbulence (see Arbic and Flierl \cite{arbicflierl03} and references therein). The relationship is seen at all depths, with potential temperature (gray shades) following the branching of the distribution. In the case with surface intensified stratification and $S=-0.6$, the presence of short-wave instabilities destroys the "sinh-like" relationship and shows a transition between local dispersive turbulence in the upper layer and non-local dispersive turbulence in the lower layer. In the upper layer (Figure \ref{f3a}c) signatures of isolated coherent structures, with distinct potential temperature signals, appear as branches separating from the cloud of points at the center of the $\psi / q$ scatter plot. In the lower layer (Figure \ref{f3a}d), the filamentary appearance of the PV anomaly field results in a cloudy appearance of the $\psi / q$ scatter plot. Moreover, in two-dimensional turbulence, the "sinh-like" appearance of the $\psi / q$ relationship is related to the presence of states of maximal entropy \cite{robertsommeria91}. Further study is requested to understand the statistical mechanics properties of the system dominated by shortwave instabilities. 

The general picture observed in the two different set of experiments remains the same if the parameters of the system are changed. In particular, the case with equal layer depths but different reduced gravities is characterized by a linear growth rate with short-wave instabilities with a maximum peak that is smaller than the peak of the long-wave instabilities. In that case, the different fields look like the fields observed for the case with equal layer depths, with potential temperature and PV gradients present around the cores of coherent structures as well as large scale gradients. Secondary roll-up instabilities are visible along the large scale gradients. No transition between the surface and the deeper layer is visible. The case with surface intensified stratification and $F_{0}=1$ is instead characterized by a larger external Rossby radius ($k_{d}=2.9$). The general picture shows the same behavior than the case with surface intensified stratification but with a sharper transition between the surface and the deeper layer. In particular, the PV anomaly field shows isolated coherent structures at the surface and elongated filaments creating sharp fronts at depth. 

\subsubsection{Probability density function of PV anomalies}
\label{sec:PVPDFs}
\begin{figure}[t]
  \noindent\includegraphics[width=30pc,angle=0]{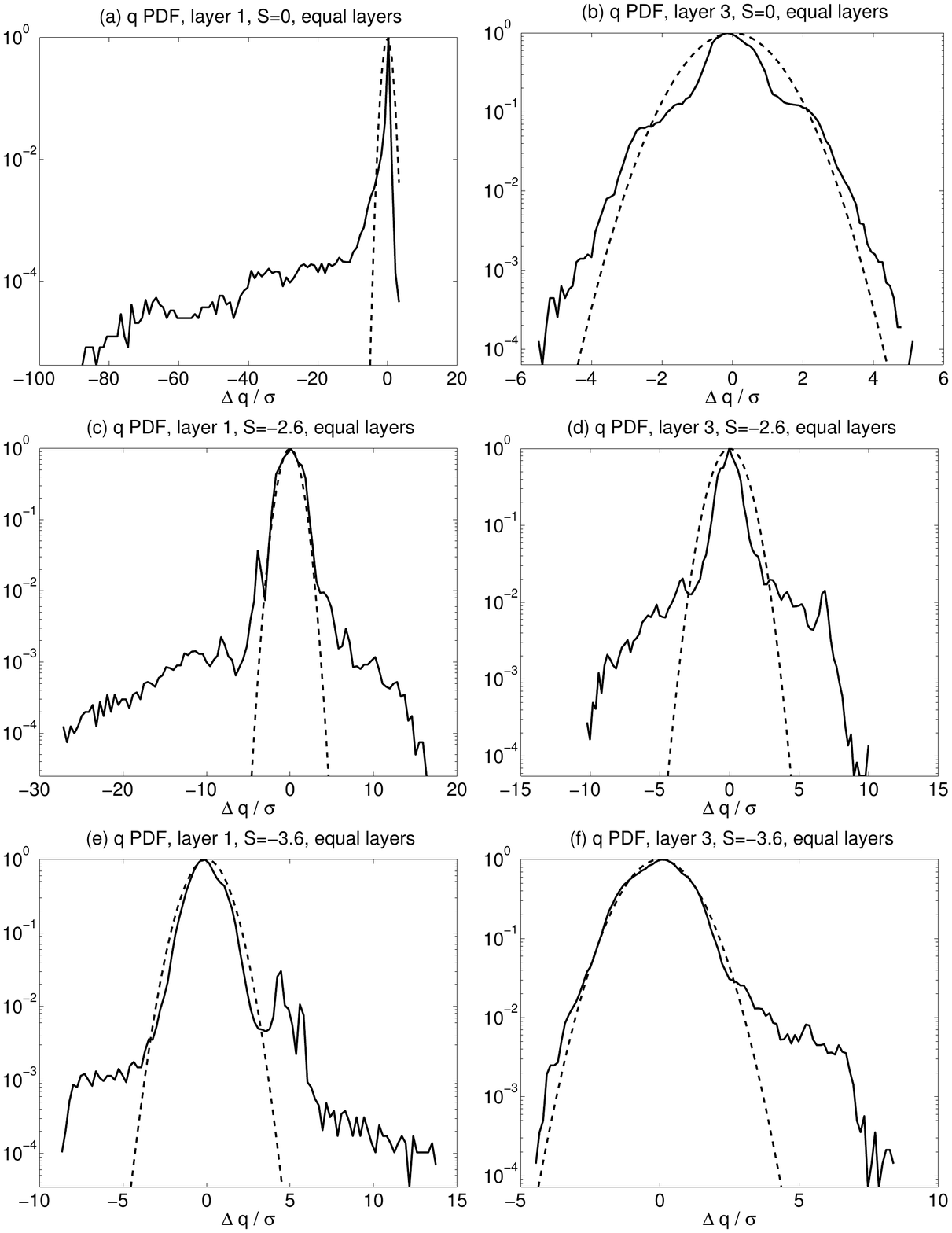}\\
  \caption{Probability density functions of the potential vorticity anomaly for layer 1 (left panels) and layer 3 (right panels) for equal layer depths and (a,b) $S=0$, (c,d) $S=-2.6$ and (e,f) $S=-3.6$. Dashed lines indicate the normal distributions calculated with the same mean and standard deviation.}\label{f5}
\end{figure}

Figure \ref{f5} shows the PDFs of the PV anomaly for the upper (left panels) and lower layer (right panels) of the system for the case with equal layer depths and for different values of the relative shear $S$, corresponding to different relative importance of short-wave instabilities. While the figures show snapshots of the PDFs, the shape remains constant in time after that the integration has reached a statistically steady state.

The results show strong, skewed deviations from the normal distributions (dashed lines) for all values of $S$ and for all depth, due to the presence of very large values of the PV anomaly in the core of the coherent structures.

\begin{figure}[t]
  \noindent\includegraphics[width=30pc,angle=0]{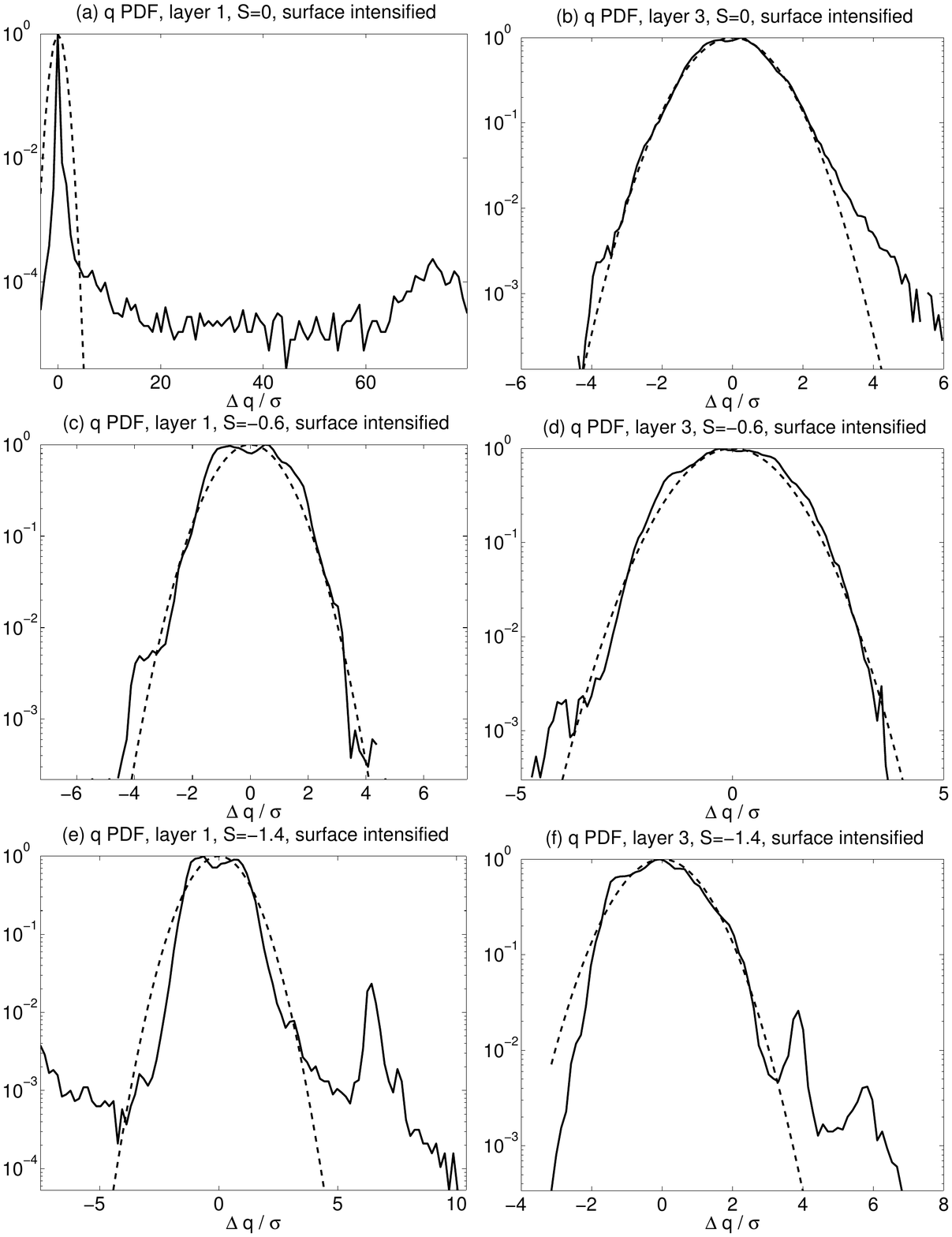}\\
  \caption{Probability density functions of the potential vorticity anomaly for layer 1 (left panels) and layer 3 (right panels) for surface intensified stratification and (a,b) $S=0$, (c,d) $S=-0.6$ and (e,f) $S=-1.4$. Dashed lines indicate the normal distributions calculated with the same mean and standard deviation.}\label{f6}
\end{figure}

A striking difference between the PDFs of the PV anomaly between the cases where the relative shear does not predict the formation of short-wave instabilities and the case in which the growth rate of the short-wave instabilities is larger than for the long-wave instabilities is however present for the case with surface intensified stratification (Figure \ref{f6}). For $S=0$ and $S=-1.4$, characterised by the absence of short-wave instabilities, the PDFs show large deviations from the normal distribution at all depths (Figure \ref{f6}a,b,e,f). However, for $S=-0.6$, correspondent to the case where the short-wave instabilities have a larger growth rate than the long-wave instabilities, the PV is redistributed to follow a Gaussian distribution at all depths (Figure \ref{f6}c,d).  

In details, the case with equal layer depths and $S=-2.6$, has skewness equal to $-2.933$ and $0.0276$ for the top and bottom layers respectively (Table 1). The same case exhibits kurtosis equal to $44.9491$ and $13.4053$ for the top and bottom layers respectively. In comparison, a normal distribution has zero skewness and kurtosis equal to 3. The case with surface intensified stratification and $S=-0.6$, corresponding to a growth of the short-wave instabilities larger than the long-wave instabilities, has skewness equal to $0.0688$ and $-0.0132$ for the top and bottom layers respectively. The same case exhibits kurtosis equal to $2.6148$ and $2.6664$ for the top and bottom layers respectively. The presence of short-wave instabilities moves thus the PDFs of the PV closer to a normal distribution, while the presence of strong coherent structures is responsible for a non-normal distribution of the PV.

\subsubsection{Vertical structure of the zonally averaged meridional eddy fluxes of PV}
\label{sec:beta}
The fact that short-wave instabilities are confined to the interface between the middle and the lower layer, introduces the question about the vertical structure of the zonally averaged meridional PV fluxes in equal layer depths and surface intensified stratification. The problem was already studied in a multi-layer system by Smith and Vallis \cite{smithvallis02}. The zonally averaged meridional PV fluxes in a three-layer system can be written as
\begin{equation}
\overline{{v'}_i {q'}_i}=-\overline{\psi_{i} \sum_{j=1}^{3} A_{i,j} \frac{\partial \psi_{j}}{\partial x}},
\label{eq:sumvq}
\end{equation} 
where the zonal divergence of the momentum term does not appear due to the setting of the problem in a double-periodic domain. Because
\begin{equation}
\sum_{i=1}^{3} H_{i} \overline{{v'}_i {q'}_i}=0,
\label{eq:sumvq}
\end{equation} 
where the over-bar indicates the zonal average, the vertical structure of the zonally averaged meridional PV fluxes can be described in a simple way using the parameter 
\begin{equation}
r=-\frac{H_{1} \overline{{v'}_1 {q'}_1}}{H_{3} \overline{{v'}_3 {q'}_3}}.
\label{eq:r}
\end{equation} 
Notice that the parameter is defined as the inverse of the same parameter defined by Held and O'Brien \cite{heldobrien92}, with negative sign retained in order to have a positive slope for $S<0$.
Assuming that the zonally averaged meridional PV fluxes act against the mean PV gradient in a downgradient diffusion with constant diffusivity $\kappa$ 
\begin{equation}
\overline{{v'}_i {q'}_i}=-\kappa \frac{\partial \Pi_{i}}{\partial y} ,
\label{eq:vqdg}
\end{equation}
where $\partial \Pi_{i} / \partial y$ is defined by (\ref{eq:pvgrad}), it is possible to write
\begin{equation}
r=-\frac{H_{1}}{H_{3}} \frac{\beta -F_{1} \left( 1 - U_{1} \right)}{\beta - F_{3}}=-\frac{H_{1}}{H_{3}} \frac{1 +\xi}{1 - \frac{F_{1}}{\beta}},
\label{eq:r2}
\end{equation} 
where 
\begin{equation}
\xi=\frac{F_{1}}{\beta} S,
\label{eq:xi}
\end{equation}
is the criticality parameter \cite{stone78}. Notice that since the problem is homogeneous, double-periodic and with a uniform zonal shear, $\kappa$ must be uniform. Notice that, as $S$ can take negative values, in the case $\xi=-1$ there is a cancellation between the thermal component of the basic PV gradient and the planetary vorticity gradient, corresponding to a baroclinic adjustment of the system \cite{stone78}. In the configuration used for this study, $\beta / F_{1} \approx O(10^{-1})$, moving the system away from marginal stability as $|S| \approx O(1)$. Notice that in this model the system can move away from criticality only locally, as the background PV gradient and the vertical shear are fixed. 

Using the identities 
\begin{equation}
\begin{array}{cc}
F_{1}=\frac{1 + 2 \epsilon}{\epsilon}F_{0}, & F_{3}=\left(1 + 2 \epsilon \right) F_{0},
\end{array}
\label{eq:nF}
\end{equation}
assuming for simplicity $\beta=1$, and assuming a PV downgradient diffusion form for the zonally averaged meridional PV fluxes, yields
\begin{equation}
r=-\epsilon \frac{1+\frac{1 + 2 \epsilon}{\epsilon}F_{0}S}{1-\left(1 + 2 \epsilon \right) F_{0}}.
\label{eq:rdg}
\end{equation} 
For $|S|>>1$, (\ref{eq:rdg}) yields the approximated ration between the slopes of $r$ against $S$
\begin{equation}
\left( \frac{\Delta r}{\Delta S} \right)_{\epsilon}= \frac{1}{1-\frac{1}{(1+2 \epsilon)F_{0}}}.
\label{eq:runi}
\end{equation} 
It is possible to see that for the choice $F_{0}=24$ and equal layer depths, $\epsilon=1$, $F_{1}=F_{3}=3 F_{0}$ and $(\Delta r / \Delta S)_{1} \approx 1.014$. For the case of vanishing upper layers, $(\Delta r / \Delta S)_{0} \approx 1.044$.

\begin{figure}[t]
  \noindent\includegraphics[width=30pc,angle=0]{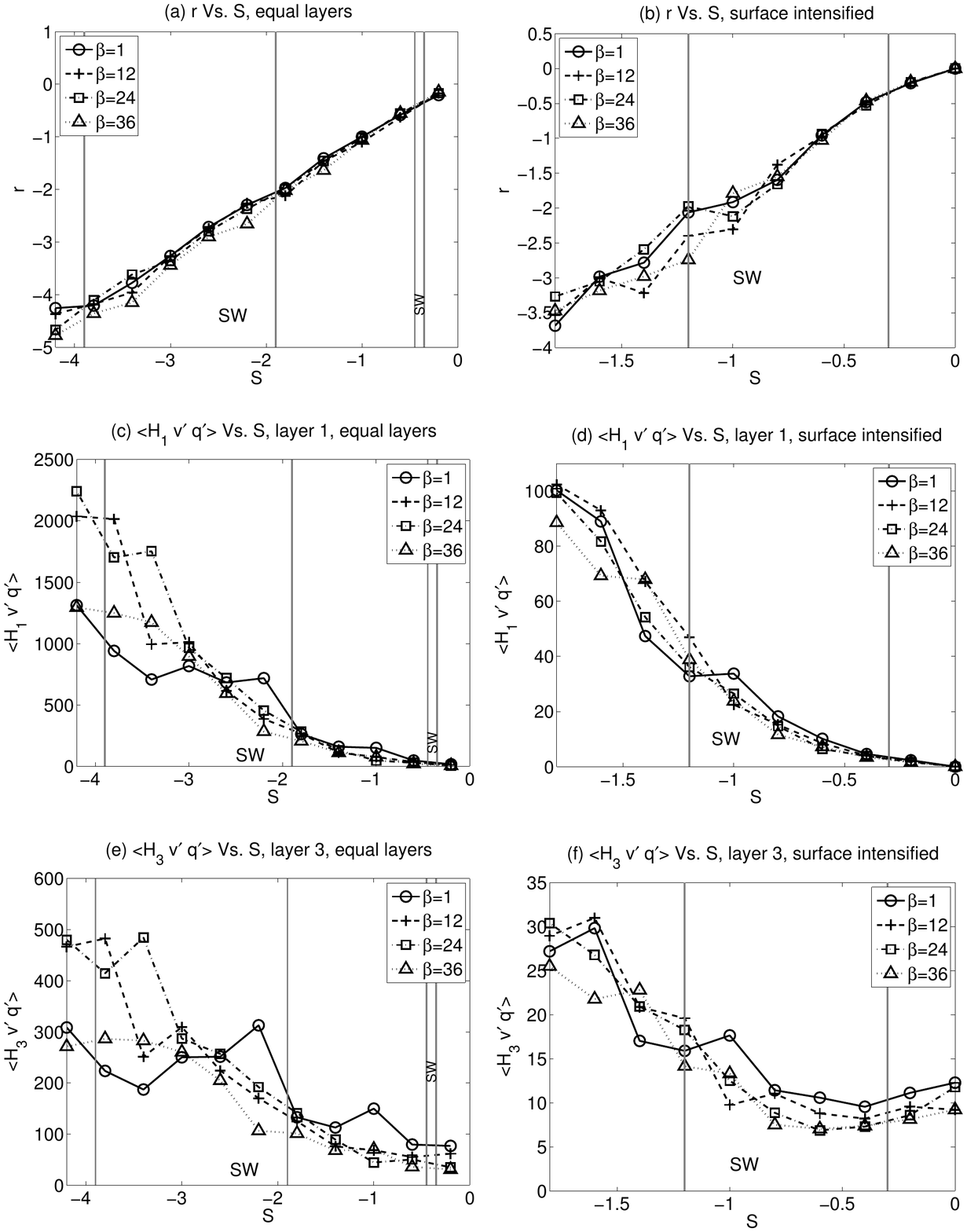}\\
  \caption{Dependence of $r$, negative of the ratio $r$ between the PV fluxes in layer 1 and layer 3, on $S$ and $\beta$ for (a) equal layer depths and (b) surface intensified stratification, for $\beta=1$ (circles, full lines), $\beta=12$ (crosses, dashed lines), $\beta=24$ (squares, dot-dashed lines) and $\beta=36$ (triangles, dotted lines). Vertical gray lines indicate the limits of the $S$ intervals inside which short-waves instabilities are present.}\label{f7}
\end{figure}

The relationship between $r$ and the relative shear $S$ is shown in Figure \ref{f7}a,b for the cases with equal layer depths and surface intensified stratification respectively. The ratio between the fluxes was calculated for different values of the non-dimensional parameter $\beta$. 

Results show that, for the case with equal layer depths, for all values of $\beta$, the parameter $r$ shows a linear relationship with the relative shear $S$ with slope $\Delta r / \Delta S \approx 1$, in agreement with the scaling law (\ref{eq:runi}). While fluctuations between the curves corresponding to different values of $\beta$ are present for values of $S$ that are far from the zero value, the trend does not show dependence on $\beta$.

The case with surface intensified stratification differs from the case with equal layer depth for a larger depth dependence of the meridional PV fluxes. The parameter $r$ shows a change of slope from a shallow slope for $S>-0.5$ to a steeper slope for $S \le -0.5$. For $S \le -0.5$. Using  the relationship between $r$ and $S$ shows a slope $1.5 < \left( \Delta r / \Delta S \right)_{0.056} < 2$. Using the value of $\epsilon=0.056$, that was used in the numerical integrations, would yield $(\Delta r / \Delta S)_{0.056} = 1.039$. The large difference between the predicted value of the slope of $r$ versus $S$ and the observed value suggests the presence of intermittency responsible for anomalous scaling. 

The larger slope for the case with surface intensified stratification can be either given by larger fluxes in the upper layer or by a different vertical structure in the fluxes resulting from the uncloupling between layers for short-wave instabilities for $S \le -0.5$, resulting in larger PV fluxes in the upper layer due to the presence of long-wave instabilities. A comparison between the magnitude of the fluxes in the upper and lower layer for the two different cases shows that the case with surface intensified stratification has inhibited fluxes in all layers (Figure \ref{f7}c-f). The larger slope of $r$ versus $S$ must thus correspond to an expected different vertical structure of the fluxes rather than to larger fluxes in the surface layer. 

In the case with equal layer depths, the size of the fluxes shows no monotonic dependence on $\beta$ (Figure \ref{f7}c,d). The case with surface intensified stratification also does not show any discernible dependence on $\beta$. Fluctuations between different curves corresponding to different values of $\beta$ are much larger than for the case with equal layer depths despite the smaller values of $S$ considered (Figure \ref{f7}e,f).

\subsubsection{Passive tracer}
\label{sec:tracer}
To assess the importance of short-wave instabilities on the lateral mixing of properties, a simple integration has been made in which a passive tracer was released. Following Smith et al \cite{smithetal02} the tracer variance is created by a fixed large scale meridional gradient, so that the advection-diffusion equation can be written as
\begin{equation}
\frac{\partial c}{\partial t} + J(\psi, c)  = - \frac{\partial \bar{c}}{\partial y} \frac{\partial \psi}{\partial x} + D_{c} ,
\label{eq:ad}
\end{equation} 
where $c$ is the tracer anomaly concentration. The first term on the r.h.s. represents the meridional advection of the mean tracer gradient, where the overbar represents the mean tracer concentration, and $D_{c}$ is the small-scale dissipation. (\ref{eq:ad}) can be rescaled so that $\partial \bar{c} / \partial y =1$. The tracer is initialized after that the model has reached statistically steady-state conditions. To focus on the lateral mixing, the tracer equation does not include vertical diffusion, that could instead be important in real geophysical fluids for its role in competing with lateral strain and thus shaping the tracer filaments \cite{haynesanglade97,smithferrari09}. Details of the numerical integration of the tracer as well as of the form of dissipation are reported in Smith et al. \cite{smithetal02}. 

\begin{figure}[t]
  \noindent\includegraphics[width=30pc,angle=0]{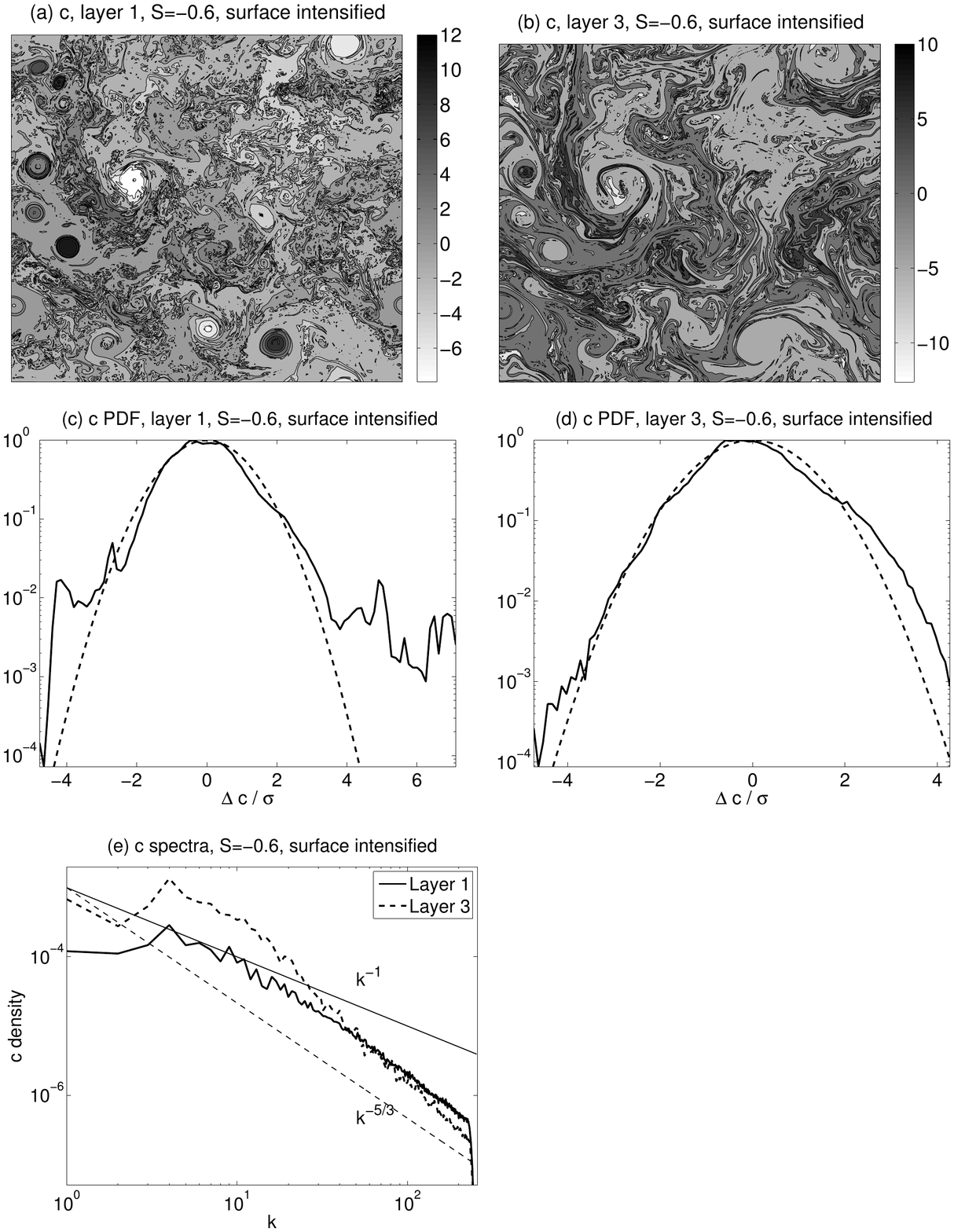}\\
  \caption{Snapshots of tracer concentration in (a) layer 1 and (b) layer 3 for surface intensified stratification and $S=-0.6$ and probability density functions for the tracer concentration for (c) layer 1 and (d) layer 3. Dashed lines indicate the normal distributions calculated with the same mean and standard deviation. (e) Tracer concentration spectra for layer 1 (full line) and layer 3 (dashed line). Thin full and dashed lines represent, respectively, lines with $-1$ and $-5/3$ slopes.}\label{f8}
\end{figure}

Snapshots of the tracer distribution are shown in Figure \ref{f8}a,b for the upper and lower layer, respectively, for the case with surface intensified stratification and $S=-0.6$, i.e. the value of $S$ for which the short-wave instabilities have a higher growth rate than the long-wave instabilities. Results show that in the upper layer the tracer is trapped in the core of the coherent structures, with stirring by small scale filaments in between them. In the lower layer part of the tracer is trapped in the cores of the coherent structures, but most of it exhibits a filamentary appearance. The change of the appearance of tracer concentration with depth is reflected in the tracer spectra (Figure \ref{f8}e). For both the surface and the bottom layer spectra (thick full and dashed lines respectively), the spectra follows a $k^{-1}$ slope at scales larger than the deformation radius, showing the local effect of the large scale coherent structures. At smaller scales, the spectra fall to a steeper slope of $k^{-5/3}$, or even steeper for the bottom layer, correspondent to non-local dispersive turbulence driven by the stirring created by the short-wave instabilities. In comparison, SQG dynamics show a complete transition from local dispersive turbulence at the surface and nonlocal at the bottom \cite{smithetal02,scott06,sukhatmesmith09} without showing a scale separation of the spectra slopes.

Finally, the role of the coherent structures in dominating the tracer dispersion is seen also in the PDFs of the tracer concentration (Figure \ref{f8}c,d, Table 1) that show large deviations from the normal distribution at the surface, with skewness equal to 0.743 and kurtosis equal to 7.1336. At the bottom, the PDF shows still deviations from the normal distribution, but the value of the skewness is 0.3595, approximately half the value of the skewness for the top layer, while the kurtosis has a value of 3.855, which is close to the value of 3, that would be expected for a normal distribution.  

\section{Discussion}
\label{sec:discussion}
An analysis of the possibility to represent short-wave instabilities using a three-layer QG model has been conducted. 

While some analogies are present between the case with surface intensified stratification and the results obtained making use of the SQG approximation, some differences can be found. As in SQG, the flow shows a transition, at least qualitiatively, between surface dynamics and interior dynamics. In particular, the case with $F_{0}=1$ (not shown) shows a clear transition between isolated coherent structures at the surface, that are deformed in elongated vortex sheets in the interior. In SQG dynamics, the presence of hyperbolic saddles in the flow can allow for singular behavior \cite{constantinetal94}, a careful analysis should thus be done to underline similarities and differences with the transition to strong fronts here reported.  

All the KE spectra show slopes different from the KE spectra in SQG, where a flattening to $k^{-5/3}$ is observed near the boundary at high wave numbers \cite{tullochsmith09a}. A SQG spectral behaviour is not expected due to the fact that the SQG modes decay in the vertical with e-folding decay proportional to the wavenumber and a high vertical resolution is required to resolve the $k^{-5/3}$ flattening of the spectra at high wavenumbers \cite{tullochsmith09}. 

The observation that in the modal KE spectra for surface intensified stratification high horizontal wavenumbers project on high baroclinic modes is due to the fact that, in a three-layer system, for shortwave instabilities the dynamics in different layers are decoupled \cite{davey77}. Numerical experiments of QG turbulence with a larger number of vertical layers, inspired by high-resolution observations of layering around mesoscale ocean eddies, show a projection of high horizontal wavenumbers on high vertical modes \cite{huaetal13}.

The analysis of the lateral dispersion of a passive tracer shows that part of the tracer is trapped within coherent structures while other parts are stirred by small scale filaments. Analysis of the tracer spectra show a transition of the spectra slopes, which suggests a scale separation of the dynamics acting on the tracer distribution. This result is different from the results obtained from SQG, where the tracer spectra show the same slope at all scales and the change from local to nonlocal dispersion is driven by the passage from surface to interior dynamics. Deviations from Gaussian distribution for both potential temperature and passive tracers are observed also in SQG dynamics at all depths \cite{scott06}, although previous arguments found here suggest a different nature of the mechanisms of intermittency acting to create anomalous scaling.

When both long and short wave instabilities are present in the linear stability analysis, at large scales the tracer seems to be primarily affected by the coherent structures. The PDFs of the tracer concentration are non-Gaussian at all depths. This has consequences on the representation of tracer stirring. For example, Haynes and Anglade \cite{haynesanglade97} use a Orstein-Uhlenbeck process to represent tracer stirring in quasi-horizontal flows that instead depends on the Gaussianity of the tracer distribution.

One of the most surprising results of this study is maybe the deviation of the PV fluxes from the fluxes obtained from a scaling that makes use of a downgradient closure. This result is is in agreement with the results obtained for the two-layer model by Vallis \cite{vallis88}. 

While the case with equal layer depths is reminiscent of observations in the Gulf Stream region, characterized by a strong PV signal at depth, results with surface intensified stratification are reminiscent of the results obtained from the linear stability analysis of oceanic configurations consisting of a mixed layer overlying a stratified interior, which result in a slow growth of interior modes and a fast growth of surface modes \cite{boccalettietal07}. Similar behavior can be obtained using the QG approximation imposing a tilt in the vertical shear and in the stratification near the surface, a configuration that is observed in so called "mixed-shear" flows, where the term "mixed" refers to a mixed westerly and easterly sheared flows, that are present in gyre-return flows at 20${}^{o}$-30${}^{o}$ of latitude \cite{tullochetal11}. However, both these configurations give rise to short-wave instabilities that are trapped at the surface, while the short-wave instabilities emerging from the three-layer model are confined to the lower interface. The system might find application in the stability analysis of currents characterized by interior PV gradients, that are not represented in the SQG approximation, such as marine undercurrents such as the ones that are formed, for example, off the California Current System. It should however be noted that, while some of the results here obtained could be used to study ocean flows, the model here used make use of a number of approximations and a lot of caution must be used before to apply it to the real world.

Finally, it should be noticed that this study focuses on the possibility to represent short-wave instabilities making use of dynamics in QG balance. It is however possible that short-wave instabilities of ageostrophic nature are created at the boundaries and can thus propagate in the interior of the ocean, as shown in both theoretical \cite{badin13} and numerical \cite{badinetal11} studies, as well as in observations \cite{calliesferrari13}. Further work is thus necessary to study the effect of ageostrophic turbulence in the interior, generated by boundary processes.



%
%

\begin{acknowledgments}
The author would like to thank F. Crisciani for fruitful discussions on the stability of the three-layer problem, D. Domeisen and A. Gabrielski for reading an early draft of the manuscript, K.S. Smith for making the numerical model available and two anonymous referees for helping improve the manuscript. The author would also like to thank the Fields Institute For Research in Mathematical Sciences, Toronto, Canada, for hospitality during the Thematic program on Mathematics Of the Ocean, in June 2013, during which part of the work presented here was first formulated. \\
\\
Appeared as:\\
G. Badin, 2014: "On the role of non-uniform stratification and short-wave instabilities in three-layer quasi-geostrophic turbulence", Physics of Fluids, 26, 096603, doi: 10.1063/1.4895590
\end{acknowledgments}


%

%
\clearpage
 \begin{table}
 \caption{\label{t1} Skewness and Kurtosis of the PDFs of the PV anomaly $q$ and tracer concentration $c$ for the values of relative shear $S$ corresponding to the exponential growth of shortwave instabilities, for the equal layer depths and surface intensified stratification}
 \begin{tabular}{l | c  c}
 {} & Skewness & Kurtosis \\
\hline \hline
q, equal layers, S=-2.6, layer 1 & -2.933 & 44.9491 \\ 
q, equal layers, S=-2.6, layer 3 & 0.0276 & 13.4053 \\ 
\hline
q, surface intensified, S=-0.6, layer 1 & 0.0688 & 2.6148 \\ 
q, surface intensified, S=-0.6, layer 3 & -0.0132 & 2.6664 \\
\hline \hline
c, surface intensified, S=-0.6, layer 1 & 0.743 & 7.1336 \\ 
c, surface intensified, S=-0.6, layer 3 & 0.3595 & 3.855 \\
 \end{tabular}
 \end{table}
 
\clearpage

\clearpage

\clearpage

\clearpage

\clearpage

\clearpage

\clearpage

\clearpage

\clearpage

\clearpage

\clearpage

\end{document}